\documentclass[aps,prl,twocolumn,showpacs,preprintnumbers,amsmath,amssymb]{revtex4}
\usepackage{epsfig}
\begin{document}
\title{Phase transitions in social networks}
\author{Piotr Fronczak, Agata Fronczak and Janusz A. Ho\l yst}
\affiliation{Faculty of Physics and Center of Excellence for
Complex Systems Research, Warsaw University of Technology,
Koszykowa 75, PL-00-662 Warsaw, Poland}
\date{\today}

\begin{abstract}
We study a model of network with clustering and desired node
degree. The original purpose of the model was to describe optimal
structures of scientific collaboration in the European Union. The
model belongs to the family of exponential random graphs. We show
by numerical simulations and analytical considerations how a very
simple Hamiltonian can lead to surprisingly complicated and
eventful phase diagram.
\end{abstract}  \maketitle

\section{I. Introduction}
During the last years, there has been noticed a significant
interest in the field of complex networks and a lot of
interdisciplinary initiatives have been taken aiming at
investigations of these systems \cite{0a,0b,0c,0d}. In parallel
with empirical studies of real world networks \cite{fal,barber},
theoretical models \cite{bar,bog,smn} and abstractive mathematical
tools \cite{new2001,fron} have been developed in order to
understand complex mechanisms hidden behind the network
functionality.

Among many studied network models like random graphs \cite{erdos},
or growing networks \cite{bar} there exists a class of models,
called exponential random graphs, which has attracted an attention
of the physics community just recently. The class was first
studied in the 1980s by Holland and Leinhardt \cite{smn9}, and
later has been developed extensively by Strauss and others
\cite{smn11,smn12,smn14,smn15}. The idea diffused from social
statistics communities to physical society in recent years, when a
number of physicists have made theoretical studies of specific
models belonging to this family
\cite{smn,smn18,smn19,smn20,smn21}.

Exponential random graph model is defined to be not a single
network but a set of possible networks (ensemble). The probability
of a particular graph $G$ in this ensemble is proportional to
$e^{-H(G)}$, where
\begin{equation}\label{eq:hamiltonian_ogolny}
H(G)=\sum_i \theta_i m_i (G),
\end{equation}
and $H(G)$ is called graph Hamiltonian, ${m_i}$ is a collection of
graph observables that reflect relevant constraints on studied
graph properties and ${\theta_i}$ is the set of fields conjugate
to ${m_i}$.

A variety of graph Hamiltonians has been studied so far including
simple random graphs \cite{smn}, a network with reciprocity
\cite{smn}, the so-called two-star model \cite{smn21}, Strauss's
model of a network with clustering \cite{solution}, and others.
Theoretical analysis of exponential random graph models has been
developed by a number of authors. In most of cases linear models
can be solved exactly in the limit of large system size. For
nonlinear Hamiltonians mean-field theory and perturbation theory
\cite{smn19,smn20, smn} have been applied in order to find phase
transitions in the network structures.

In this paper we would like to show how a very simple Hamiltonian
can lead to surprisingly complicated and eventful phase diagram
where wealth of structural phase transitions can not be forecast
at first glance. Due to complexity of observed structures our
methodology is mostly concentrated on Monte Carlo simulations. A
simple mathematical apparatus is also expounded in order to reveal
details of the observed phenomena. The mentioned calculations,
although not so powerful, allow to understand when and why a
particular transition occurs.

\section{II. Motivation and model description}

The model is defined on network that is composed of $N$ nodes and
$L$ links, where each node acts as a single scientist or a single
scientific group and a link between two nodes means that there
exist scientific collaboration between them.

The original purpose of the model was to describe optimal
structures of scientific collaboration in the Sixth Framework
Programme for Research and Technological Development (FP6), which
is the European Union action aiming to  stimulate and support
scientific activities conducted at national and international
level. One of main purposes of the European Commission financing
scientific projects in FP6 was to strengthen co-operation between
project partners \cite{FP6}. In their proposal, applicants had to
show that one of the aims of the planned project is
intensification of co-operation between participants. They also
had to argue that without such an interaction a goal of the
project will be not achieved. The most popular observable which
allows to measure effects of the co-operation in social network is
{\em clustering coefficient} introduced by Watts and Strogatz in
1998 \cite{Watts}. The clustering coefficient $c_i$ of a single
node $i$ is the proportion of the number of links between the
nodes within its neighborhood $e$ divided by the number of links
that could possibly exist in the neighborhood
\begin{equation}
c_i=\frac{2 e}{k_i (k_i-1)},
\end{equation}
where $k_i$ represent degree of the considered node. If $k_i <2$
then $c_i = 0$. The global clustering coefficient $C$ is just an
average of $c_i$ over all nodes.

The other obvious purpose of the funded project is to achieve the
highest possible productivity. In our model productivity of each
scientist $i$ (or local scientific group) depends on the number of
collaborators $k_i$. The more collaborators work with a given
scientist, the more papers/ideas the scientist can produce. On the
other hand, however, a large number of collaborators means the
necessity of parallel concentration on different scientific
threads which leads to the decrease of productivity. In
consequence, productivity $p_i$ of a single project participant
can be modelled by a logistic-like equation
\begin{equation}
p_i=k_i(1-\frac{k_i}{h})\stackrel{k_i<h}{\approx} k_i e^{-k_i /
h},
\end{equation}
where $h$ is an optimal (desired) number of collaborators (note
that the highest productivity occurs for $k_i=h$). Although one
could expect that in reality $h$ should be described by a kind of
the Lotka distribution \cite{Lotka}, here we concentrate on the
simplest case where $h$ is the same for all elements of the
system. Productivity $P$ of the whole network is just an average
over all $N$ nodes normalized to unity
\begin{equation}
P=\frac{e}{N h} \sum_i p_i.
\end{equation}

Hamiltonian of the described model can be written as follows:
\begin{equation}
H(G)=-\theta P(G)-\alpha C(G).
\label{H}
\end{equation}

Monte Carlo procedure is defined by the following algorithm: we
choose randomly two nodes and try remove (add) existing (non
existing) link between them. If the change leads to the decrease
of the initial system energy $E_0$, i.e. $\Delta E=E_{m/p}-E_0<0$,
where $E_{m/p}$ is the system energy after link removal/addition,
we accept such a replacement. Otherwise, when $\Delta E\geq 0$, we
accept it with the probability $e^{-\Delta E}$, i.e. we apply
typical Metropolis algorithm.

\section{III. Analytical considerations}

In order to show a large variety of structural transitions
observed in networks described by the Hamiltonian (\ref{H}), in
our Monte Carlo simulations we have decided to fix one parameter
$\alpha=10^9$, and check behavior of the system as a function of
the second parameter $\theta$. At the moment please note, that
since one may think of parameters $\alpha$ and $\theta$ as inverse
temperatures our choice of the large value of $\alpha$ corresponds
to low temperature limit in classical thermodynamics, and makes
the considered system less susceptible to random effects. Further
in the paper, the assumption of large $\alpha$ allows us to
estimate the critical value of $\theta$ from the simple condition
$\Delta E=0$, which means that the stability of the initial stable
structure is no longer preserved and a new network configuration
can emerge.

\begin{figure*}[!ht]
 \centerline{\epsfig{file=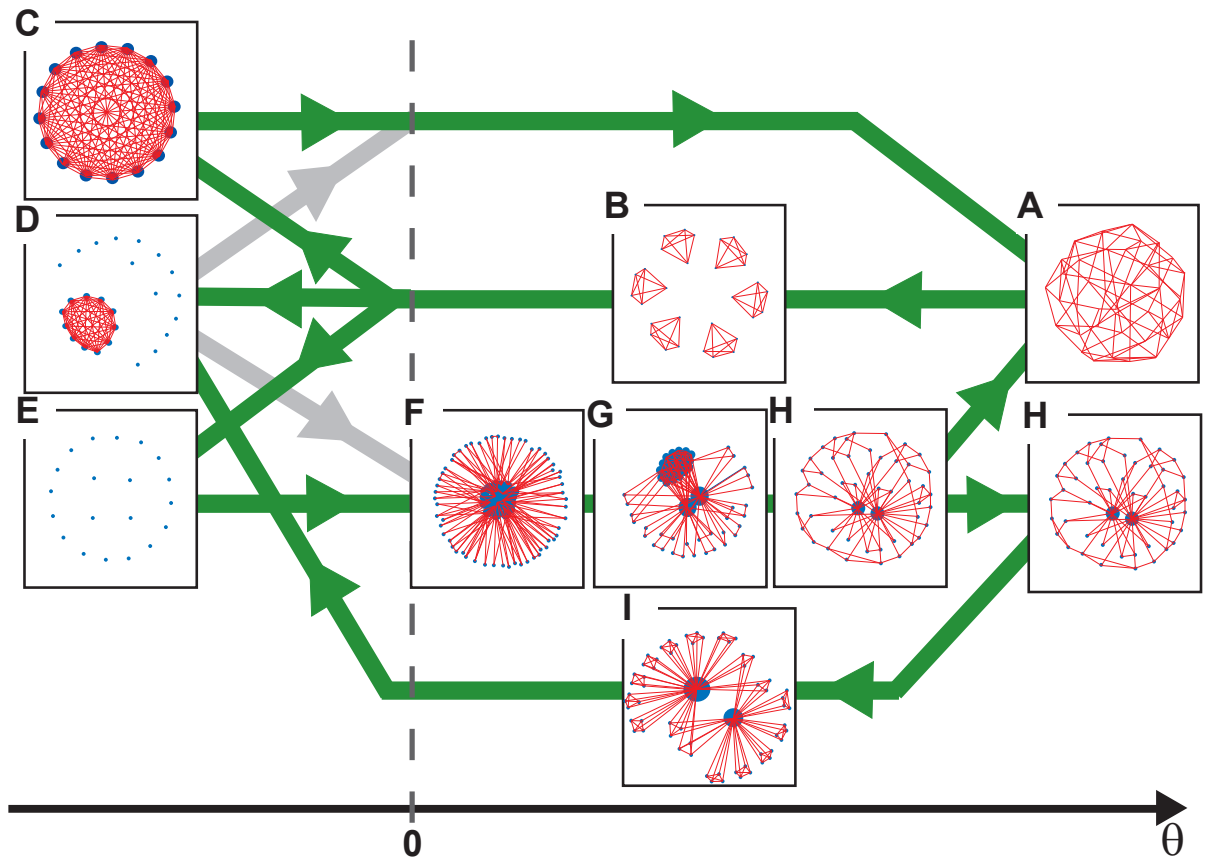,width=16cm}}
    \caption{(Color online) Diagram representing possible transitions
    between network configurations as a function of parameter $\theta$. Gray arrows means that
    the configuration $D$ is composed of two other configurations ($C$ and $E$), and the two parts
    of the system follow different paths. The figure shows networks with
    different sizes $N$ just to emphasize particular character of
    a given configuration.}
 \label{fig:diagram}
\end{figure*}

Fig. \ref{fig:diagram} shows possible scenarios of structural
transitions in our model. Arrows represent directions of changes
of the control parameter $\theta$ given $\alpha=10^9$. As one can
see the number of possible transition paths is impressive. Later
we show that the path chosen by the system depends mainly on the
network size $N$. Moreover, beside simple paths like $ABCA$ or
$ABEFGHA$ there may exist much more complicated paths like
$ABEFGHIDFGHA$, i.e. we have to change parameter $\theta$ from
$+\infty$ to $-\infty$ and backward from $-\infty$ to $+\infty$
two or more times to return to the same structure we started!

\begin{figure}
 \centerline{\epsfig{file=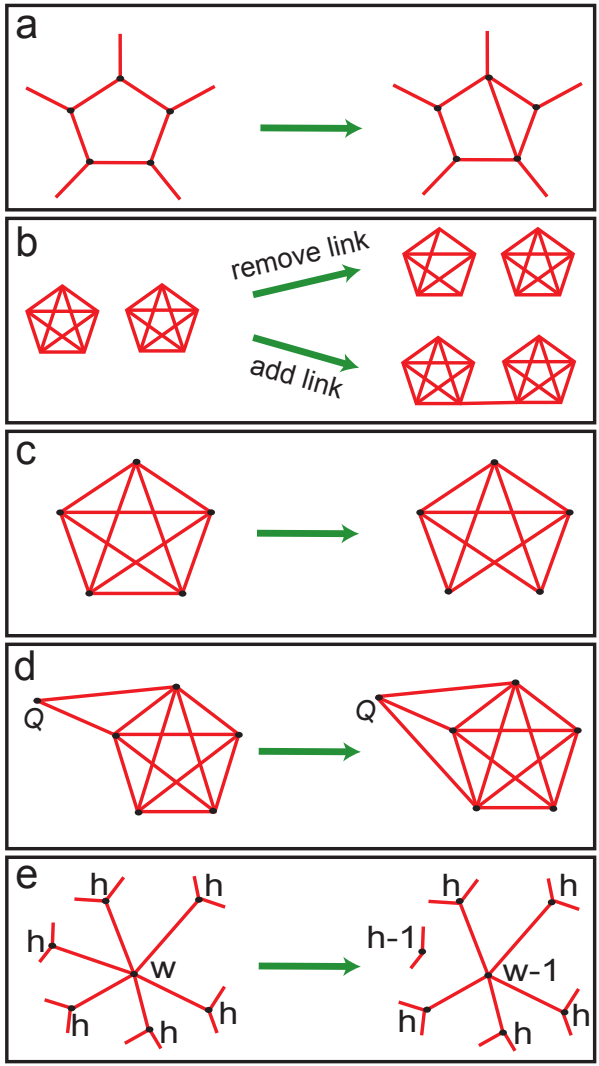,width=8cm}}
    \caption{(Color online) Schematic situations occurring during
    particular transitions used for considerations performed in the text. a)
    transition $A-B$; b) transition $B-D$; c) transition $C-A$;
    d) transition $E-F$; e) transition $H-A$.}
 \label{fig:schematic}
\end{figure}

To find critical value of the parameter $\theta$ at which a
particular structural transition occurs one has to analyze change
in energy $\Delta E$ induced by link addition or removal taking
into consideration currently existing network structure. As an
example let us analyze transition $A-B$. The structure $A$
corresponds to regular random graph, where node degree
distribution is given by the delta function $P(k)=\delta(k-h)$. In
this structure productivity is maximal $P=1$, whereas the number
of triangles contributing to the clustering coefficient is
negligible small $C\simeq 0$ (assuming that graph is sparse i.e.
$h \ll N$ that is sociologically justified). The transition takes
place when for a particular value of the parameter $\theta$
energetically favorable is to add a link which creates the first
triangle (i.e. a decrease of productivity is sufficiently rewarded
with an increase of clustering). The described situation is
schematically presented in Fig. \ref{fig:schematic}a. Energies
corresponding to both structures depicted in the figure are
respectively given by
\begin{equation}
\begin{array}{ll}
  E_0 = & -\theta \cdot 1-\alpha \cdot 0, \\
  E_p = & -\theta\left[(N-2)h e^{-1}+2(h+1) e^{-\frac{h+1}{h}}\right]\frac{e}{N h} \\
   & -\alpha\left[\frac{4}{N h(h+1)}+\frac{2}{N h (h-1)}\right].  \\
\end{array}
\end{equation}
First, inserting the values of energy into the condition $\Delta
E=E_p - E_0 =0$, and next expanding exponential functions in
Taylor series up to the second order one gets the critical value
of the control parameter for the considered transition $A-B$
\begin{equation}
\theta_{A-B}\approx\frac{2(h+1)(3h-1)\alpha}{h^2}.
\end{equation}
As one can see the transition does not depend on the system size $N$.

\begin{figure}
 \centerline{\epsfig{file=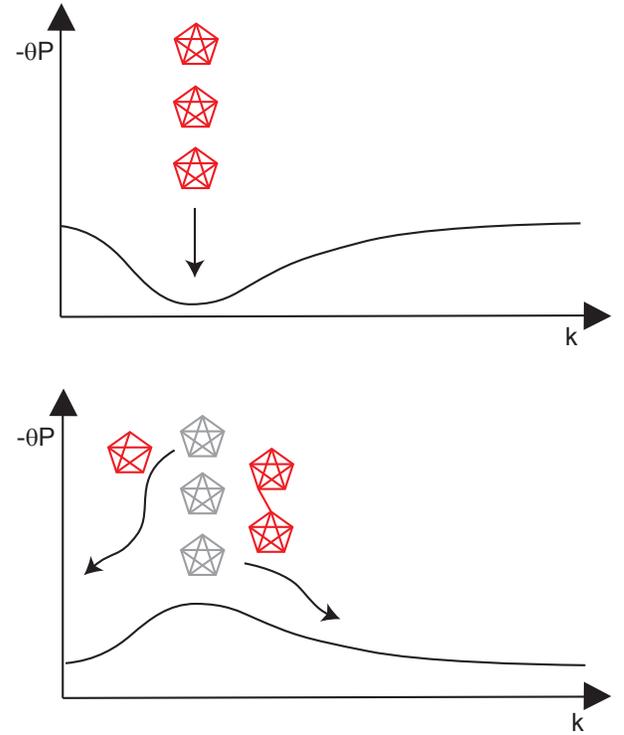,width=8cm}}
    \caption{(Color online) Productivity contribution to energy for
    $\theta>0$ (upper figure) and for $\theta<0$ (lower figure) as
    a function of node degree $k$. The later situation shows that
    to decrease energy some nodes will reduce their degree
    (destroying existing clusters) and others will increase it
    (connecting two clusters together).}
 \label{fig:nonstable}
\end{figure}

Much more complicated system behavior is observed when the control
parameter $\theta$ crosses zero and becomes negative (see
Fig.~\ref{fig:diagram}). Productivity contribution to energy
changes sign and all nodes having at the moment degree $k=h$ turn
out to be in unstable configuration (see schematic explanation
given in Fig.~\ref{fig:nonstable}). For such nodes when decrease
in clustering is sufficiently rewarded by decrease in undesirable
productivity the stable configuration $B$ will be destroyed: some
nodes will decrease their degrees whereas others will increase
them (cf. Fig.~\ref{fig:schematic}b). As one can see in
Fig.~\ref{fig:diagram} the considered network may follow one of
three paths resulting in one of three configurations $C$, $D$, or
$E$.

\begin{figure}
 \centerline{\epsfig{file=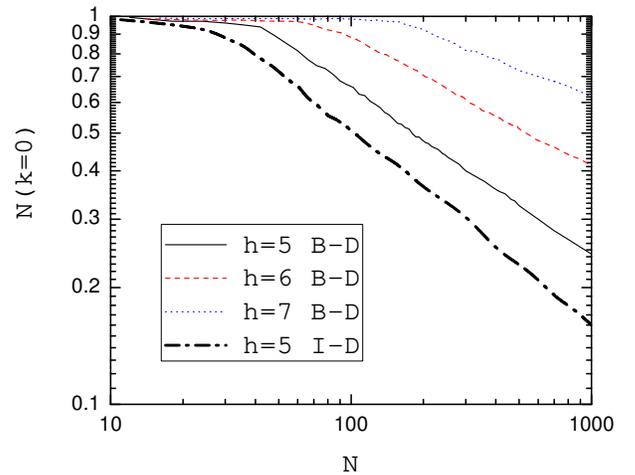,width=8cm}}
    \caption{Number of nodes with degree $k=0$ as a function of system size $N$.
    Three thin curves represent a transition $B-D$, while a thick one represents
    a transition $I-D$. The thick curve shows that for the same value of the parameter
    $h$ the number of isolated nodes is lower after transition $I-D$
    than after transition $B-D$. }
 \label{fig:isolated}
\end{figure}

Unfortunately, due to probabilistic character of the Monte Carlo
procedure it is hard to calculate analytically which direction of
changes will be taken by the system of a given size $N$. To check
what is really happened during the transition we perform numerical
simulations, results of which are summarized in Fig.
\ref{fig:isolated}. The figure shows a fraction of isolated nodes
as a function of $N$. As one can see for small system sizes
degrees of all nodes drop to zero and the system transforms into
the empty graph (configuration $E$ in Fig. \ref{fig:diagram}).
Above some critical value of $N$ a part of nodes condensate
together and a fully connected subgraph accompanied by isolated
nodes appears (configuration $D$ in Fig. \ref{fig:diagram}).
Finally in the thermodynamical limit all nodes condensate and the
complete graph emerges (configuration $C$ in Fig.
\ref{fig:diagram}).

Here, we have to stress that configurations $C$ and $E$ are rather
purely theoretical and can not appear in real systems. In the case
$E$, it is connected with the fact that in the configuration $B$ a
small number of nodes with degree larger than $h$ can exist what
helps to create nodes with maximal degree during the transition.
Nevertheless, since the configuration $D$ is in fact composed of
the two remaining configurations $C$ and $E$, it is still
instructive to analyze them.

Now, let us analyze the transition $C-A$, i.e. transition from the
complete graph to the regular random graph. Because one can not
add a link to the complete graph the only situation to analyze is
removal of a link. The described situation is schematically
presented in Fig.~\ref{fig:schematic}c. A simple calculation gives
\begin{equation}
\begin{array}{ll}
  E_0 = & -\theta N(N-1)e^{-\frac{N-1}{h}}\frac{e}{N h}-\alpha, \\
  E_m = & -\theta\left[(N-2)(N-1)e^{-\frac{N-1}{h}}+2(N-2)e^{-\frac{N-2}{h}} \right]\frac{e}{N h} \\
   & -\alpha\left[1-\frac{2}{N(N-1)}\right],  \\
\end{array}
\end{equation}
which leads to
\begin{equation}
\theta_{C-A}\approx\frac{e^{N/h}h^2\alpha}{N^2}
\end{equation}
in the limit of large $N$. As one can see this transition depends
on the system size $N$, and in the thermodynamical limit
$N\rightarrow\infty$ the critical value $\theta_{C-A}$ tends to
infinity.

Now, let us discuss the behavior of the system if the initial
configuration is $E$ (i.e. the empty network). The first
transition $E-F$ occurs when the parameter $\theta$ equals zero.
Since the productivity does not influence energy of the system at
this point, links can appear randomly (they do not increase energy
so they are acceptable in the Monte Carlo procedure). First
triangles appear in the network and the clustering coefficient
increases. Such a dynamics leads to the configuration in which
fully connected subgraph is surrounded by a number of peripheral
nodes with degree $k=2$ (configuration $F$ in Fig.
\ref{fig:diagram}). Because of complicated situations in
intermediate time steps a rigorous analytical explanation of the
transition $E-F$ is beyond our abilities. Nevertheless, below we
analyze a simplified situation, which allows one to understand why
the transition occurs.

\begin{figure}
 \centerline{\epsfig{file=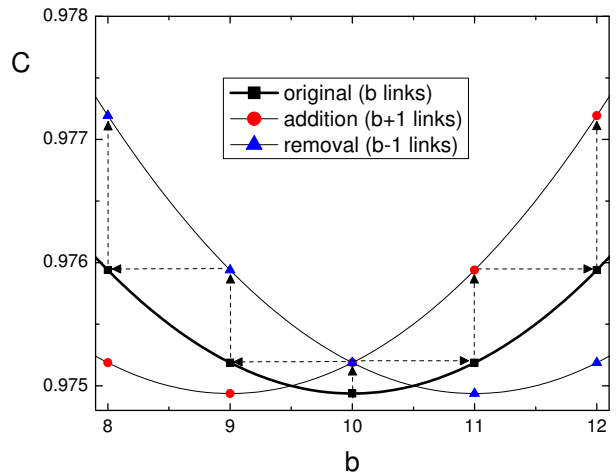,width=8cm}}
    \caption{Clustering coefficient $C$ for the system depicted in Fig.~\ref{fig:schematic}d. Arrows show evolution of
    the system (addition or removal of successive links).}
 \label{fig:b_node}
\end{figure}

Thus, let us consider a fully connected subgraph with an
additional node $Q$ having $b$ links (see
Fig.~\ref{fig:schematic}d). Just like before, one can analyze what
happens with the clustering coefficient if we add (remove) one
link. Fig.~\ref{fig:b_node} shows the solution of this problem
when size of the fully connected subgraph is $N_F=20$. For a given
$b$ clustering coefficient of the considered structure is marked
by the thick line. Thin lines show a new clustering coefficient
after addition (circles) or removal (triangles) of one link. At
the beginning, let us assume that we have $b<10$. To increase
clustering coefficient we have to remove one link which leads to
the configuration with $b-1$ links attached to the peripheral node
$Q$. Further, it is easy to see that the process will follow
towards removing next links belonging to $Q$ until the node will
have only two links. The node degree can not drop below $k_Q=2$,
because for $k_Q<2$ local clustering coefficient $c_Q$ suddenly
drops from $1$ to $0$ which drastically decreases global
clustering coefficient of the whole structure. On the other hand,
if we assume that $b>10$, then energetically favorable is to add
another $b+1$ link to the node $Q$. Again, one can see that the
node $Q$ will try to connect to all other nodes, i.e. the node $Q$
becomes a member of the fully connected subgraph.

Starting from the configuration $F$ if one further increases
$\theta$ the productivity $P$ starts to matter. It means that
above some critical value of this parameter an addition of the
third link to one out of peripheral nodes with degree $k=2$ can
compensate energy lost coming from decrease of the clustering
coefficient. The same reasoning explains successive network
reorganizations when adding next links (up to $k=h$) to peripheral
nodes is energetically favorable.

\begin{figure}
 \centerline{\epsfig{file=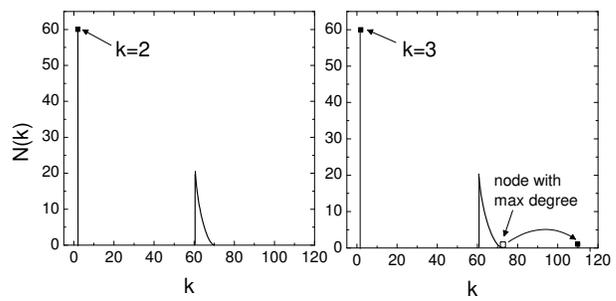,width=8cm}}
    \caption{Schematic node degree distribution before and after transition $F-G$.
    The node with the highest degree is exaggerated to show its dynamics during the transition.}
 \label{fig:maxdegree}
\end{figure}

At the moment, let us note that nodes belonging to the fully
connected subgraph in the configuration $F$ have different
degrees. Their degrees are composed of $N_F-1$ mutual links and
links coming from peripheral nodes with degree $k=2$ (the
peripheral links are randomly distributed among the nodes creating
the fully connected subgraph). It means that in a finite system
there always exists a node with the largest degree. It is easy to
check that addition of a new link to this node makes clustering
coefficient of the whole structure worse in the least way. It
means that above a certain value of $\theta$ new links are added
to this node making its degree rapidly growing.
Fig.~\ref{fig:maxdegree} shows in a schematic way the node degree
distribution below and at the critical value of $\theta$ for the
transition.

As we have described above, if one further increases the control
parameter $\theta$ the peripheral nodes suddenly increase their
degrees from $k=3$ to $k=4$. It generates a similar mechanism as
described in the previous paragraph, i.e. consecutive node with
the largest degree (expect the one which has already increased its
degree to $N$) stepwise increases its connectivity. The network
configuration arising along the transition corresponds to the
configuration $G$ presented in Fig.~\ref{fig:diagram}.

Another crucial point in the system evolution is the transition
$G-H$, where the fully connected subgraph is destroyed in the
similar manner as the complete graph in the transition $C-A$.
After the transition $G-H$ our network consists of several hubs
and a large number of loosely connected peripheral nodes.

\begin{figure}
 \centerline{\epsfig{file=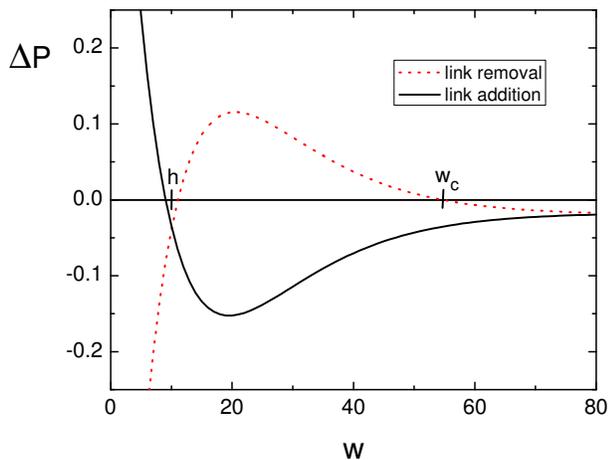,width=8cm}}
    \caption{Change of productivity in the star-like structure depicted in Fig. \ref{fig:schematic}e
    as a function of degree of the central highly connected node $w$, for $h=10$.}
 \label{fig:w}
\end{figure}

The configuration $H$ is presented twice in Fig. \ref{fig:diagram}
in order to show two possibilities of the system evolution: it can
be stable when $\theta\rightarrow\infty$, or hubs can be
destroyed, and the transition $H-A$ takes place. To analyze
stability of the configuration $H$ is enough to consider a
simplified structure presented in Fig.~\ref{fig:schematic}e (note
that the simplified star-like structure neglects effects of
clustering which may occur in the original configuration $H$). In
such a structure, change of productivity resulting from addition
or removal of a single link between the hub with a large degree
$w$ and one of peripheral nodes with degree $h$ (as we have
already stated the maximal connectivity of the peripheral nodes
along the transition path $F-G-H$ is $k=h$) is presented in
Fig.~\ref{fig:w}
\begin{equation}
\begin{array}{ll}
  P_0 = & \left[ he^{-1}+we^{-\frac{w}{h}}\right]\frac{e}{N h}, \\
  P_m = & \left[ (h-1)e^{-\frac{h-1}{h}}+(w-1)e^{-\frac{w-1}{h}}\right]\frac{e}{N h}, \\
  P_p = & \left[ (h+1)e^{-\frac{h+1}{h}}+(w+1)e^{-\frac{w+1}{h}}\right]\frac{e}{N h}. \\
\end{array}
\end{equation}
A given process (link addition or removal) occurs only if the
change of productivity is positive. It means that if $w$ is small
the only possibility is to add a new links. When $h<w<w_{c}$ (c.f.
Fig.~\ref{fig:w}) links can only be removed. It means that degree
of the hub should decrease from $w$ to $h$, and the transition
$H-A$ takes place. On the other hand, when $w>w_c$, both processes
(addition and removal of links) are no longer possible. It means
that the system remains in the stable configuration $H$.
Unfortunately, because the analysis neglects clustering it does
not allow us to calculate the precise critical value of
$\theta_{H-A}$.

Finally, let us analyze the transition $H-I$
(Fig.~\ref{fig:diagram}), which occurs when starting from the
stable configuration $H$ one decreases the control parameter
$\theta$. On the basis of our previous considerations one can
predict that the configuration $I$ emerges when the input of
clustering coefficient to the system energy crosses a critical
value. In a similar way like during the transition $A-B$ nodes
with low degrees have tendency to form triangles. On the other
hand, however, because hubs are stable (we have shown it in the
previous paragraph) instead of many separated clusters like in the
configuration $B$ the system evolves towards the stable
configuration $I$.

If the parameter $\theta$ is sufficiently negative the
configuration $I$ is destroyed just like the configuration $B$,
and the mechanism of this transition is the same: which
configuration (C,D, or E) appears depends on the system size N
(c.f. thick dotted curve in Fig.~\ref{fig:isolated}). The only
difference is that for a given $N$ the number of isolated nodes is
lower in comparison with the transition $B-D$.

As we have stated before, in fact, our system in the configuration
$D$ consists of two configurations $C$ and $E$. It means that at
least to some value of the parameter $\theta$ the part of the
system that is equivalent to configuration $E$ follows a path
$D-F-G-H$, and the second part, equivalent to the configuration
$C$, follows the path $C-A$. At some point, up to now separated
parts of the network combine together. In figure
\ref{fig:diagram}, in order to make the whole picture as clear as
possible, we marked paths accessible for these two components of
the configuration $D$ by gray arrows.

Let us also notice that a path our system follows can be really
complicated. For example, let us consider a network in the
configuration $A$. After transitions $A-B-D$ the number of
isolated nodes is high, which allows to create hubs with very high
degrees $w>w_{c}$ as a result of the transition $F-G$. It means
that the configuration $H$ is stable, and the system follows the
path $H-I-D$. Now, the number of isolated nodes is much lower (see
Fig. \ref{fig:isolated}). Their number is often too small to once
more time create hubs with degrees $w>w_{c}$ during the transition
$F-G$. Thus, after a series of transitions $D-F-G-H$ the system
returns to the initial state $A$. It means, that the return to the
same state is possible after a complicated path
$A-B-D-F-G-H-I-D-F-G-H-A$, in which the chain of transitions
$D-F-G-H-I$ may be repeated several times.

Finally, we have to stress that a part of phase transitions we
observed are visible only in finite size systems, i.e. if the
system is large enough a particular phase transition can change
its character and can lead to development of different structure.
Although usually physicists use term {\em phase transition} in the
context of systems in thermodynamical limit, our delinquency can
be justified because the model we study has been proposed to
social systems where the number of elements is always limited.

\section{IV. DISCUSSION}

As we have presented in the previous section our simple model is
characterized by a surprisingly complicated and eventful phase
diagram with plenty of metastable states. Nevertheless, since the
model was sociologically motivated, let us discuss the observed
network configurations in the context of scientific collaboration.

First, the configuration $B$ seems to be the easiest to interpret.
If each node represents scientific group then we see here
separated projects consisting of several scientific groups where
each group collaborate with each other. The realism of the
situation can be questionable since real projects can be composed
of different number of participants, but let us remind that in the
model we have assumed that the optimal number of collaborators $h$
is fixed for all groups. To make the model more realistic we
should draw the parameter from the Lotka-like distribution
\cite{Lotka}, but it would certainly complicate obtained results
and at the moment we were much more interested in the description
of the observed structural phase transitions.

\begin{figure}
 \centerline{\epsfig{file=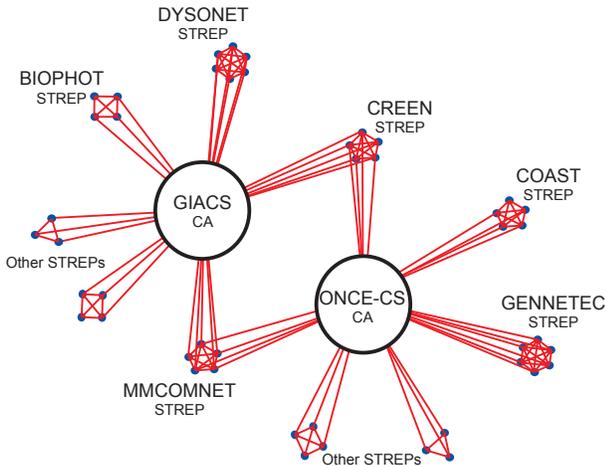,width=8cm}}
    \caption{(Color online) Schematic view on network of EU projects.
    Two highly connected coordination actions are surrounded by a plenty of small research projects.}
 \label{fig:giacs}
\end{figure}

The second configuration which seems to reflect real-world
observations is the configuration $I$. Fig.~\ref{fig:giacs} shows
the real (although simplified) case of EU projects being currently
in progress. Let us explain that complex systems research in
Europe is funded through two European Commission actions: NEST
(New and Emerging Science and Technology, and FET-IST (Future and
Emerging Technology – Information Society Technology) \cite{FP6}.
It is mainly being done through small projects called STREPS
(Strategic Targeted Research Projects). Apart from them the
European Commission is currently funding two Coordination Actions
to support complex systems science: ONCE-CS (funded by IST-FET)
and GIACS (funded by NEST). In Fig.~\ref{fig:giacs} they are
presented as main nodes which serve as knowledge transfer units
between particular projects. Part of projects funded mainly by
NEST collaborate with GIACS, and projects funded mainly by IST-FET
are supported by ONCE-CS. Of course there are projects which take
support from both coordination actions because aims of both CA's
are slightly different. Let us stress that the above picture is
very simplified. The main simplification is that CA's are
represented by single nodes in Fig.~\ref{fig:giacs}, which is
evidently not true - coordination actions are projects consisting
of many participants as well as STREPs.

Finally, the last comment. A careful reader can ask what could be
the interpretation of negative value of the control parameter
$\theta$. In fact, it corresponds to the situation where groups
composed of small number of participants are undesirable. The
sociological explanation (although not connected with scientific
collaboration) can be expressed as follows: {\em or you commune
with one global social group or you will be separated}, which can
be recognized as the fascist ideology.

\section{V. CONCLUSIONS}

In this paper we have presented a model of social collaboration.
Although the model is expressed by a simple Hamiltonian the
richness of observed structural phase transitions is impressive.
Most of them we can only analyze qualitatively and further studies
have to be performed to clarify reasons for which a given
structure appears. We uncover many aspects of the studied model
but in fact much more questions arise during our investigations.

Although simplifications of the model do not allow to render in
detail the real-world space of scientific projects, we have shown
that some configurations formed in the system remind existing
structures of European projects.

\section{Acknowledgments}

Authors wish to thank Krzysztof Suchecki and Julian Sienkiewicz
for fruitful discussions and insightful comments. JH acknowledges
a financial support from the State Committee for Scientific
Research in Poland (Grant No. 134/E-365/6.PR UE/DIE239/2005-2007).
P.F. acknowledges a support from the EU Grant CREEN
FP6-2003-NEST-Path-012864. A.F. acknowledges financial support
from the Foundation for Polish Science (FNP 2006).

\end{document}